\documentclass{iopjournal}

\usepackage{graphicx}
\usepackage{cite}
\usepackage{url}
\usepackage[dvipsnames]{xcolor}
\usepackage{amsmath,amssymb,bigints,multirow}
\usepackage[numbers,sort&compress]{natbib}

\makeatletter
\renewcommand\@seccntformat[1]{\csname the#1\endcsname.\quad}
\makeatother

\begin{document}

\articletype{Research article}

\title{Matter- and magnetically-driven flavor conversion of neutrinos in magnetorotational collapses}

\author{Marco Manno$^{1,2,3,*}$~\orcid{0009-0009-8622-8594}, Pablo Mart\'inez-Mirav\'e$^3$~\orcid{0000-0001-8649-0546} and Irene Tamborra$^3$~\orcid{0000-0001-7449-104X}}

\affil{$^1$Dipartimento di Matematica e Fisica ``Ennio De Giorgi,'' Universit\`a del Salento, 73100, Lecce, Italy}

\affil{$^2$INFN-Istituto Nazionale di Fisica Nucleare, Sezione di Lecce, 73100, Lecce, Italy}

\affil{$^3$Niels Bohr International Academy and DARK, Niels Bohr Institute, University of Copenhagen,

Blegdamsvej 17, 2100, Copenhagen, Denmark}

\noindent $^*$Corresponding author.

\email{marco.manno@unisalento.it, pablo.mirave@nbi.ku.dk, tamborra@nbi.ku.dk}

\begin{abstract}
The magnetorotational collapse of massive stars copiously emits neutrinos of all flavors, with a prominent hierarchy between the non-electron and electron flavor average energies. Relying on a three-dimensional  neutrino-magnetohydrodynamic simulation of a $13 M_\odot$ progenitor, we investigate flavor conversion in matter. We find that, in addition to resonant flavor conversion of neutrinos and antineutrinos in matter, (anti)neutrinos  experience  chirality-flipping interactions due to their non-zero magnetic moment  ($\mu \lesssim 10^{-12} \mu_B$) and  large magnetic field in the source ($B \simeq 10^{15}$~G). For Majorana neutrinos, this leads to resonant flavor-changing neutrino-antineutrino mixing. The event rate expected from  a Galactic collapse at current and next-generation neutrino telescopes, such as IceCube and Hyper-Kamiokande, strongly depends on the  orientation of the magnetorotational collapse with respect to the observer direction and flavor conversion scenario. The event rate is expected to be larger for an observer facing head on the jet launched during the stellar collapse and peaks around $400$--$600$~ms  after bounce. Our work highlights that understanding the rich phenomenology of flavor conversion in magnetorotational collapses is essential to take full advantage of the joint detection of neutrinos and gravitational waves from these sources.  
\end{abstract}

\section{Introduction}

At the end of their lives, stars with mass  $M\gtrsim 8M_\odot$ undergo  gravitational collapse: a core-collapse supernova (CCSN). The delayed  neutrino-driven explosion mechanism   is believed to drive the majority of stellar core collapses~\cite{Bethe:1985sox}--see, e.g., Refs.~\cite{Janka:2012wk,Muller:2020ard,Burrows:2020qrp} for reviews on the topic. Up to a few percent of stellar collapses is characterized by rapid rotation and large magnetic fields. In this case, the progenitor angular momentum is high enough to spin up the proto-neutron star.
The  strong differential rotation provides a substantial reservoir of energy; even relatively weak pre-collapse magnetic fields  undergo  amplification through rotational winding and, more efficiently, by means of magnetorotational instabilities achieving toroidal field strengths of $B \simeq 10^{15}$--$10^{16}$~G. In such cases, the stellar collapse is magnetorotationally   driven~\cite{BisnovatyiKogan1970,LeBlanc:1970kg,Meier1976,Wheeler2002,BalbusHawley1991,Akiyama2003}. Magnetorotational collapses are often invoked  to explain hyper-energetic explosions, such as superluminous supernovae and a subset of gamma-ray bursts~\cite{Aguilera-Dena:2018ork,Obergaulinger:2021omt}.

Similar to  neutrino-driven explosions~\cite{Mirizzi:2015eza,Raffelt:2025wty,Tamborra:2024fcd}, copious emission of neutrinos is expected from collapsars~\cite{Just:2022fbf,Nagataki:2007xm,Harikae:2009dz,Obergaulinger:2017qno,Fujibayashi:2022xsm,Issa:2024sts,Shibata:2025gix,Dean:2024hhu,Fernandez:2025qbq}; however, larger average
energies for the non-electron flavors are expected in magnetically-driven collapses  due to the fact that neutrino decoupling takes place at baryon densities larger than for neutrino-driven
 supernovae~\cite{Martinez-Mirave:2024zck}. Neutrinos from magnetorotational collapses are  expected to contribute to the diffuse supernova neutrino background, adding to  the cumulative emission coming from neutrino-driven collapses, neutrino-driven accretion flows and accretion disks~\cite{Kresse:2020nto,Moller:2018kpn,Ando:2023fcc,Schilbach:2018bsg,Wei:2024qgy,Martinez-Mirave:2024zck,Martinez-Mirave:2025pnz}. 

In order to  forecast the joint  detection prospects of neutrino and gravitational waves from nearby collapses of rotating massive stars~\cite{Tamborra:2024fcd,Mirizzi:2015eza,Horiuchi:2018ofe}, it is essential to understand neutrino flavor evolution in the source. The latter is expected to be driven by neutrino-neutrino interactions in the innermost regions of the collapsing star~\cite{Pantaleone:1992eq,Sawyer:2008zs,Sawyer:2015dsa} and  matter effects in the less dense regions~\cite{Wolfenstein:1977ue,Mikheyev:1985zog,Dighe:1999bi}; cf.~Refs.~\cite{Mirizzi:2015eza,Tamborra:2020cul,Volpe:2023met,Johns:2025mlm} for recent reviews on the topic.

Neutrinos are also expected to have a non-zero magnetic dipole moment~\cite{Giunti:2014ixa}. Lower bounds coming from theoretical arguments  suggest $\mu \gtrsim \mathcal{O}(10^{-23}$--$10^{-22}) \mu_B$, with $\mu_B = e/(2 m_e)$ being the Bohr magneton.  The most stringent bounds from terrestrial experiments are compatible with $\lesssim \mathcal{O}(10^{-11}) \mu_B$, while  astrophysical bounds give $\lesssim \mathcal{O}(10^{-12}) \mu_B$, see Refs.~\cite{Giunti:2014ixa,ParticleDataGroup:2024cfk} and references therein. Hence, in the presence of large magnetic fields, 
neutrinos may experience magnetically-induced flavor conversion~\cite{Fujikawa:1980yx,Cisneros:1971,Schechter:1981hw,Voloshin:1986ty,Okun:1986na}. If neutrinos are  Majorana particles, neutrino-antineutrino flavor conversion can take place~\cite{Akhmedov:1988uk, Akhmedov:2003fu,Abbar:2020ggq, Dvornikov:2011dv,Giunti:2014ixa}. However,  the physics of neutrino flavor conversion in magnetorotational collapses has been poorly explored. 

In this paper, we investigate  the physics of resonant  flavor conversion in matter, in the presence of large magnetic fields, focusing on Majorana neutrinos. To do so, we rely on input from the three-dimensional neutrino-magnetohydrodynamic  model  from Ref.~\cite{obergaulinger-aloy}. We also explore the  neutrino event rate expected at the IceCube Neutrino Observatory~\cite{Abbasi:2011ss} and Hyper-Kamiokande~\cite{Hyper-Kamiokande:2018ofw}. 
Because of the asymmetries developed during the magnetorotational core collapse, we find a strong variability of the expected event rate  in existing and upcoming neutrino telescopes according to the observer direction.  

Our paper is organized as follows. In Sec.~\ref{sec:mrcc}, we outline the main features of the neutrino emission properties from our  reference three-dimensional simulation. Section~\ref{sec:conversions} presents our findings on the phenomenology of resonant flavor conversion of neutrinos in matter  and neutrino-antineutrino flavor-changing  conversion in the presence of  non-zero magnetic moments. The oscillated neutrino fluxes at Earth are then explored in Sec.~\ref{sec:flux_conversion}. In Sec.~\ref{sec:detection}, we investigate the expected event rate in the IceCube Neutrino Observatory and Hyper-Kamiokande, accounting for different scenarios of flavor conversions and observer directions. Finally, our results are summarized  in Sec.~\ref{sec:conclusion}.

\section{Benchmark model}
\label{sec:mrcc}
In this section, we introduce the main features of the three-dimensional neutrino-magnetohydro\-dynamic simulation that we adopt. We also characterize the  flavor-dependent neutrino emission properties. Due to the strong directional dependence of the thermodynamic and hydrodynamic properties as well as the neutrino emission characteristics, hereafter, we focus on the equatorial and polar emission from our model (i.e., we focus on emission directions on-axis and perpendicular  to the jet launched during the core collapse); intermediate features are to be expected for intermediate emission directions.

\begin{figure*}[t]
  \centering
  \includegraphics[width=0.95\linewidth]{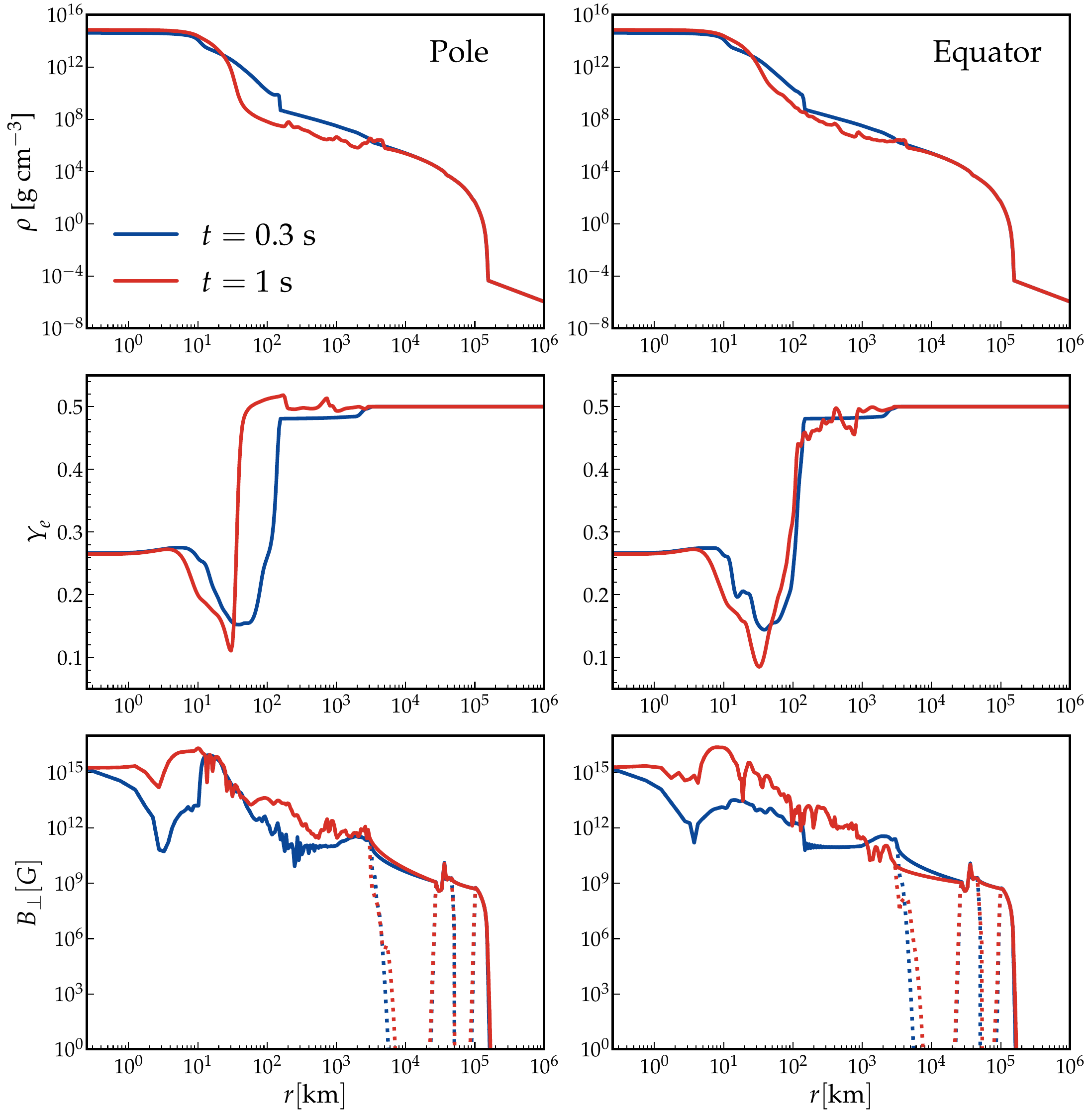}
 \caption{Radial profiles of the baryon density  ($\rho$, top row), electron fraction  ($Y_e$, middle row), and transverse magnetic-field strength ($B_\perp=\sqrt{B_\theta^2+B_\varphi^2}$, bottom row) at two post-bounce times 
  ($t=0.3\,\mathrm{s}$ and $t=1\,\mathrm{s}$), shown along the pole ($\theta=0$ and $\varphi =0$, left column) 
  and the equator ($\theta=\pi/2$ and  $\varphi =0$, right column). The dotted lines in the bottom panel represent the output of the neutrino-magnetohydrodynamic simulation, but are unphysical and due to the progenitor  model employed to initialize the  simulation. The solid lines represent our  interpolated profiles  that smoothly bridge these unphysical gaps (see main text for more details). }
  \label{fig:profiles}
\end{figure*} 
\subsection{Magnetorotational core collapse model}
We rely on a three-dimensional special relativistic neutrino-magnetohydrodynamical simulation of a progenitor with a zero-age main sequence mass of $13 M_\odot$  presented in Ref.~\cite{obergaulinger-aloy}  (corresponding to the two-dimensional model of Ref.~\cite{Obergaulinger:2021omt}). The simulation accounts for two-moment (M1) neutrino transport and  uses the SFHo hadronic equation of state~\cite{Steiner:2012rk}. 
The progenitor undergoes chemically homogeneous evolution, due to enhanced rotational mixing. According to the 2D equivalent simulation of Ref.~\cite{Obergaulinger:2021omt}, the core collapse results in a protomagnetar~\cite{Metzger:2010pp,Metzger:2018szx}; the post-collapsed core  develops a  magnetic field strength of $\gtrsim 10^{15}$~G, with core bounce occurring at  $t_b = 0.25$~s. We refer the interested reader to Refs.~\cite{obergaulinger-aloy, Obergaulinger:2021omt} for additional details on the core-collapse dynamics and explosion properties.

Figure~\ref{fig:profiles}  displays, from top to bottom, the radial profiles of the baryon density ($\rho$), the electron fraction ($Y_e$),  and the magnetic-field component perpendicular to the radial direction ($B_\perp$), for  our selected post-bounce times $t=0.3$~s and $t=1$~s. Each of these quantities  is a function of radius, the polar and azimuthal angles, and the post-bounce time; i.e., $\rho \equiv  \rho(t, r,\theta,\varphi)$, $Y_e \equiv Y_e(t, r,\theta,\varphi)$, and $B_\perp \equiv B_\perp(t, r,\theta,\varphi)$. To simplify the notation, we omit the explicit dependence on such coordinates hereafter. In order to assess the variability of these characteristic quantities according to the emission direction, we select  two representative directions employing the grid of the magnetorotational collapse simulation: the pole ($\theta = 0$ and $\varphi=0$, left column) and the equator ($\theta = \pi/2$ and $\varphi=0$, right column). Such directions are representative of the emission properties along the jet launched during the collapse and those in the plane perpendicular to the jet direction.

The baryon density profiles (top row of Fig.~\ref{fig:profiles}) show  a dense proto-neutron star core  surrounded by an extended mantle with steeply decreasing density gradient. The profiles are qualitatively similar along the two selected directions and throughout the evolution. As the post-bounce evolution proceeds from $0.3$ to $1$~s, the inner layers contract and reach higher central densities, while  the shock propagates outward. Notice that there are clear direction-dependent differences at intermediate radii ($20$--$10^3$~km), determined by the  non-spherical dynamics and magnetically-driven outflows. In particular, there is a steeper decrease in density outside the proto-magnetar in the polar direction, whereas the density decrease is smoother along the equator. Notice also that the outermost layers, for radii larger than $10^4$~km, negligibly depend on the post-bounce time and observer direction. 

The electron fraction $Y_e$ is plotted in the middle row of Fig.~\ref{fig:profiles}. In the dense proto-neutron star interior, where electron degeneracy is high and neutrinos are trapped, $Y_e \simeq 0.3$. Moving outwards, a pronounced dip in $Y_e$ develops corresponding to the deleptonization layer, produced by efficient electron capture ($e^- + p \rightarrow n + \nu_e$), which dominates neutrino emission and drives the local electron fraction to a minimum. Beyond this layer, absorption of electron neutrinos off neutrons ($\nu_e + n \rightarrow p + e^-$)  outweighs electron capture, pushing the composition toward a more symmetric proton-to-neutron ratio. As a result, $Y_e$ grows sharply at  $t=0.3$~s and $t=1$~s, along both emission directions. As visible in Fig.~1, $Y_e$ exceeds $0.5$ only along the polar direction at $t=1$~s, indicating the development of proton-rich regions.

The bottom row of Fig.~\ref{fig:profiles} represents $B_\perp$, the component of the magnetic field relevant for the 
neutrino–antineutrino conversion discussed in Sec.~\ref{sec:Bres}. The inner regions exhibit  strong  magnetic-field strength ($B_\perp \simeq 10^{15}$--$10^{16}$~G  in the proximity of the proto-magnetar), reflecting the combined effect of flux compression and rotational winding. 
At larger radii, the field gradually decreases with radius and displays significant directional variability, as visible from the differences between the polar and equatorial profiles.. At intermediate and large radii ($r \simeq 10^4$~km  and just below $10^5$~km), the dotted curves show the simulation output exhibiting localized gaps, where the magnetic field artificially drops to zero~\cite{grifiths-thesis}. These features are not  physical;  they arise from the progenitor  model (used to initialize our benchmark simulation), where the magnetic field profile is  prescribed  in radiative regions only, while convective zones have  $B=0$. As a consequence, the magnetorotational simulation retains regions of vanishing field in the vicinity of the  convective shells. Since such gaps are not expected to occur
 and would artificially affect our analysis, 
we construct interpolated profiles (solid curves) that smoothly bridge these unphysical gaps. The interpolated profiles are the ones adopted in the rest of this work.

\subsection{Neutrino emission properties}
\label{sec:neutrino_properties}
\begin{figure}
\centering
\includegraphics[width=0.95\linewidth]{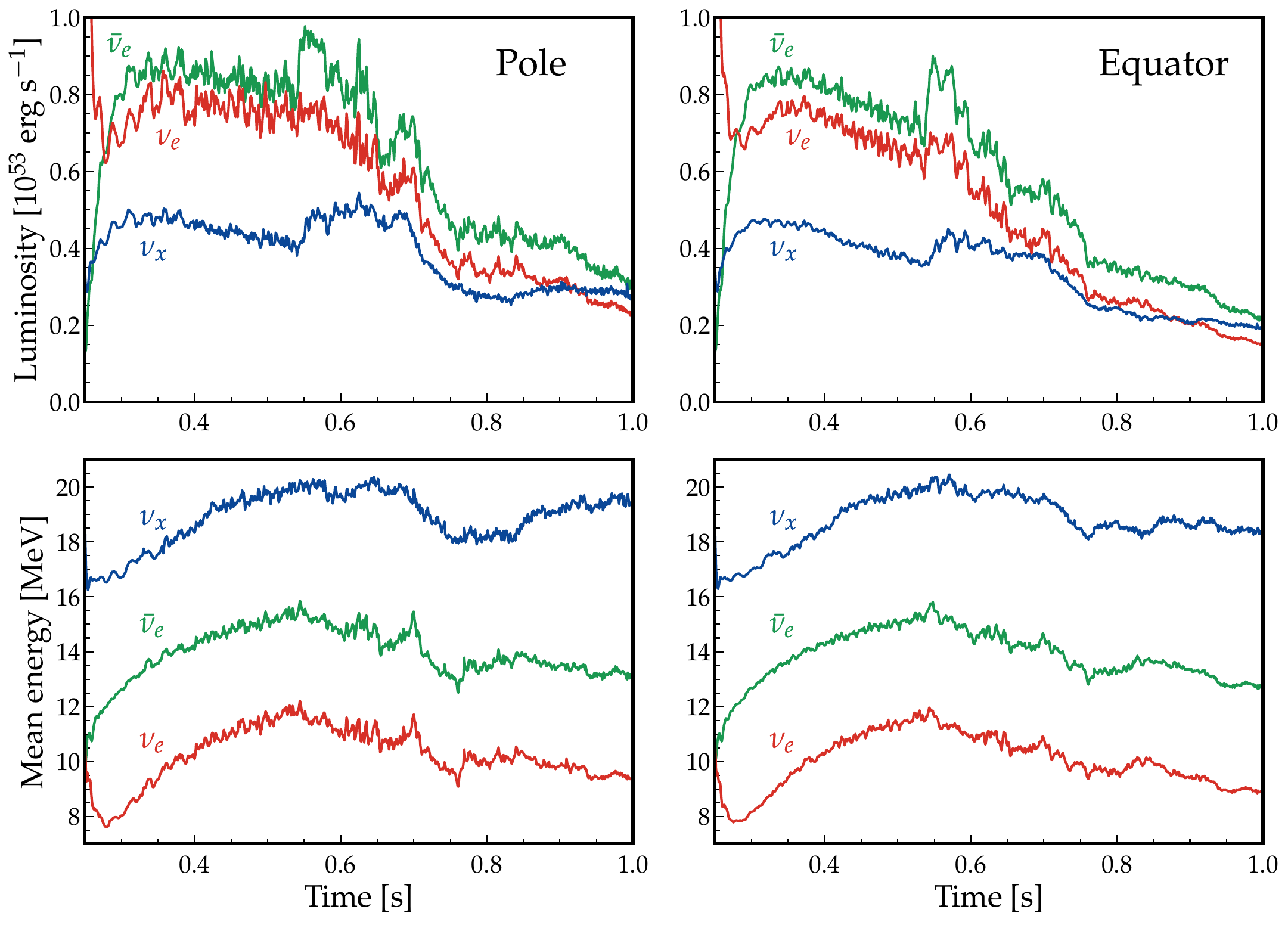}
\caption{
Time evolution of the neutrino luminosity (top row) and mean energies (bottom row) for electron neutrinos ($\nu_e$), electron antineutrinos ($\bar{\nu}_e$), and heavy-lepton species ($\nu_x=\nu_\mu$, $\nu_\tau$, $\bar{\nu}_\mu$, or $\bar{\nu}_\tau$). The left and right panels represent the neutrino emission properties as measured by a distant observer with angular coordinates close to  the pole ($\theta=0$ and $\varphi =0$) and  equator ($\theta=\pi/2$ and $\varphi = 0$), respectively.}
 \label{fig:lum_en}
\end{figure}
Figure~\ref{fig:lum_en} shows the time evolution of the luminosities (top row) and mean energies (bottom row) of electron neutrinos ($\nu_e$), electron antineutrinos ($\bar{\nu}_e$), and non-electron species ($\nu_x=\nu_\mu$, $\nu_\tau$, $\bar{\nu}_\mu$, or  $\bar{\nu}_\tau$) as seen by an observer at the north pole and the equator (with the observer coordinate system being the same as the one of the  simulation grid). The neutrino emission characteristics have been extracted at  $10^3$~km and  projected to the observer following the procedure outlined in Appendix A of Ref.~\cite{Tamborra:2014hga}. Also for the neutrino emission characteristics, we omit the explicit angular dependence for simplicity: $L \equiv L(t,\theta,\varphi)$ and $\langle E\rangle \equiv \langle E\rangle(t,\theta,\varphi)$.

Right after core bounce, rapid neutronization  results in a burst of electron neutrinos. Later, during the accretion-dominated phase, the emission is such that $L_{\bar\nu_e} > L_{\nu_e} > L_{\nu_x}$, where $x$ denotes any of the non-electron flavors ($\nu_x = \nu_\mu$, $\nu_\tau$, $\bar{\nu}_\mu$, or $\bar{\nu}_\tau$). 
Eventually, the luminosity of non-electron neutrinos overcomes the one of electron neutrinos, towards the end of the temporal interval of interest in this work (i.e.~$t\lesssim 1$~s).

As for  the mean energies,  a clear hierarchy is visible, with  $\langle E_{\nu_x}\rangle > \langle E_{\bar{\nu}_e}\rangle > \langle E_{\nu_e}\rangle$. The mean energies tend to increase between $t=0.3$~s and $0.5$~s and flatten out or slowly decrease afterwards. For orientation, the temporal averages (up to $1$~s post bounce) of the flavor-dependent mean energies are:  $\langle E_{\nu_x}\rangle \simeq 19$~MeV, $\langle E_{\bar{\nu}_e}\rangle\simeq 14$~MeV, and $\langle E_{\nu_e}\rangle\simeq 10$~MeV. Notice that the mean energy of non-electron flavor neutrinos is larger than the one characterizing  core-collapse supernovae, as highlighted in Ref.~\cite{Martinez-Mirave:2024zck}; this feature is relevant for detection purposes, especially in the presence of flavor conversion.

The temporal evolution of the luminosity and the first energy moment are  qualitatively similar along the pole and the equator (left and right panels of Fig.~\ref{fig:lum_en}, respectively). However,  the polar direction shows larger variability on small time scales. In the context of neutrino-driven collapses, it has been shown that hydrodynamical instabilities are responsible for inducing  modulations in the neutrino and gravitational wave signals; see, e.g., Refs.~\cite{Tamborra:2013laa,Tamborra:2014hga,Fernandez:2025qbq,Walk:2019miz,Walk:2018gaw,Drago:2023cve,Lund:2010kh,Vartanyan:2019ssu,Nagakura:2020qhb,Lin:2022jea}. We leave the  analysis of the physics linked to the modulations of the neutrino emission properties in magnetorotational collapses  to future work.

After accounting for weighted averages of the neutrino emission properties in the observer direction~\cite{Tamborra:2014hga}, and in the absence of flavor conversion, the differential flavor-dependent number flux of a generic flavor species $\nu_\ell$ is: 
\begin{equation}
    \Phi^0_\ell(E)
    = \frac{1}{4\pi D^2}\,
      \frac{L_\ell}{\langle E_\ell \rangle}\,
      f_\ell(E)\,,
    \qquad \text{for} \,\,
    \ell = \nu_e,\ \bar{\nu}_e,\ \nu_x\, ,
    \label{eq:flux}
\end{equation}
where $D$ is the distance of the magnetorotational collapse from Earth, and $L_\ell$ and $\langle E_\ell\rangle$ are the luminosity and the mean energy of flavor $\ell$, see Fig.~\ref{fig:lum_en}.
The normalized quasi-thermal spectrum of each flavor is approximated by a Gamma distribution~\cite{Keil:2002in,Tamborra:2012ac},
\begin{equation}
f_\ell(E)=
\frac{E^{\alpha_\ell}}
{\Gamma(\alpha_\ell+1)}
\left( \frac{\alpha_\ell+1}{\langle E_\ell\rangle} \right)^{\alpha_\ell+1}
\exp\!\left[-(\alpha_\ell+1)\frac{E}{\langle E_\ell\rangle}\right]\, ,
\end{equation}
where the pinching parameter $\alpha_\ell$ follows from the first two energy moments:
\begin{equation}
\alpha_\ell =
\frac{\langle E_\ell^2 \rangle - 2\langle E_\ell\rangle^2}
{\langle E_\ell\rangle^2 - \langle E_\ell^2 \rangle}\, .
\end{equation}

\section{Flavor  conversion in matter}\label{sec:conversions}
Neutrino flavor conversion affects the (anti)neutrino flux emitted from the source. Neutrino self-interaction is expected to be the driver of flavor change in the innermost regions of the core, where the neutrino density is the largest (neglected in this work). At larger radii, neutrino-matter interactions further affect the flavor composition. We distinguish between two different types of conversion. First, the Mikheyev-Smirnov-Wolfenstein (MSW) resonant conversion  that is driven by  neutrino-matter interactions. Second,  since  magnetorotational collapses host large magnetic fields and neutrinos are expected to have non-zero magnetic moments, flavor-changing neutrino-antineutrino conversion can further affect the flavor composition of Majorana neutrinos. 
In this section, we investigate the physics of MSW conversion and  flavor-changing neutrino-antineutrino conversion.

\subsection{Resonant  flavor conversion}
\label{sec:MSW}
The flavor evolution of neutrinos and antineutrinos can be described by the following equation 
\begin{align}
    i \frac{\rm d \varrho}{{\rm d}t} = [\mathcal{H}, \varrho]\, ,
    \label{eq:evolution_eq}
\end{align}
where $\varrho$ is the $6 \times 6$ density matrix. Its diagonal and off-diagonal elements account for the occupation number of each neutrino and antineutrino flavor and the flavor coherence of the system, respectively. The Hamiltonian in the flavor basis is
\begin{align}
\label{eq:HMSW}
    \mathcal{H} = \begin{pmatrix}
        H & \mathcal{O} \\
        \mathcal{O} & \bar{H}
    \end{pmatrix} \, ,
\end{align}
with $\mathcal{O}$ being a null matrix  and
\begin{align}
 \label{eq:hamiltonian31}
    H &= \frac{1}{2E} U\text{diag}(0, \Delta m^2_{21}, \Delta m^2_{31})U^\dagger + \text{diag}(V_e + V_n, V_n, V_{\mu\tau} +V_n)\, ,\\ \bar{H} &=\frac{1}{2E} \bar{U}\text{diag}(0, \Delta m^2_{21}, \Delta m^2_{31})\bar{U}^\dagger - \text{diag}(V_e + V_n, V_n, V_{\mu\tau} +V_n)\, .
    \label{eq:hamiltonian3}
\end{align}
The $3\times 3$ Hamiltonians, $H$ and $\bar{H}$, account for neutrino and antineutrino flavor evolution, respectively; the first term is linked to neutrino mixing in vacuum,  $E$ is the neutrino energy and $U$ is the  mixing matrix parametrized as in Ref.~\cite{ParticleDataGroup:2024cfk}, in terms of three mixing angles denoted by $\theta_{12}$, $\theta_{13}$, $\theta_{23}$, and a complex phase $\delta$. The mass splittings are $\Delta m^2_{21}$ and $\Delta m^2_{31}$, with $\Delta m^2_{ij} = m^2_j - m^2_i$. 
We define the mass-mixing parameters relying on the best-fit values provided in  Ref.~\cite{deSalas:2020pgw} in normal and inverted ordering (NO and IO, respectively), as summarized in Table~\ref{tab:osc-params}. Moreover, we use $\delta \neq 0$, although it is expected to have a negligible impact on the flavor conversion phenomenology  due to the fact that our magnetorotational model does not distinguish between mu and tau flavors~\cite{Balantekin:2007es}. For antineutrinos, $\bar{U}$ is defined as $U$ but replacing $\delta \to -\delta$, as a consequence of CP symmetry.
\begin{table}[]
    \centering
    \renewcommand{\arraystretch}{1.4}
    \caption{Neutrino mixing parameters adopted in this work for NO and IO, as from Ref.~\cite{deSalas:2020pgw}.}
    \begin{tabular}{lcccccc} \hline\hline
      &$\sin^2\theta_{12} $ & $\sin^2\theta_{13}$  & $\sin^2 \theta_{23}$ & $\delta/\pi$ & $\Delta m^2_{21} [10^{-5}$ eV$^{2}]$  & $\Delta m^2_{31}[10^{-3}$eV$^{2}]$ \\ \hline
      NO  & $0.318$ & $0.02200$ & $0.574$ & $1.08$ & $7.5$ & $2.55$ \\
      IO  & $0.318$ & $0.02225$ & $0.578$ & $1.58$ & $7.5$ & $-2.45$\\ \hline\hline
    \end{tabular}
       \label{tab:osc-params}
\end{table}

The second term of the Hamiltonian accounts for matter effects and includes the potential induced by coherent forward scattering of neutrinos off electrons and neutrons, $V_e$ and $V_n$, respectively. Such potentials can be expressed in terms of the baryon density and electron fraction as
\begin{align}
    V_e = \sqrt{2}G_F\frac{\rho Y_e}{m_N} \qquad \text{and}\qquad V_n = -\frac{1}{\sqrt{2}}G_F\frac{\rho(1-Y_e)}{m_N}\, ,
\end{align} 
where $G_F$ is the Fermi constant and $m_N$ is the mass of a nucleon. We also account for the difference in the effective potential experienced by $\nu_\mu$ and $\nu_\tau$ ($V_{\mu\tau}$),  stemming from radiative corrections~\cite{Botella:1986wy}. 

The Hamiltonians in Eqs.~\ref{eq:hamiltonian31} and \ref{eq:hamiltonian3} should also include a third term modeling neutrino self-interaction, which is, however, neglected in this work. Neutrino-neutrino interaction is expected to operate in the innermost regions of the magnetorotational collapse and may have a non-negligible impact on the flavor composition, similar to the case of core-collapse supernovae--we refer the interested reader to Refs.~\cite{Mirizzi:2015eza,Tamborra:2020cul,Volpe:2023met,Johns:2025mlm} for recent reviews on the topic. However, because of the conceptual and technical challenges  linked to  neutrino-neutrino interaction in dense media and because flavor conversion has been very poorly investigated in the context of magnetorotational collapses, we choose to focus on understanding the impact of matter-induced flavor conversion. 

Resonant flavor conversion (MSW resonance)~\cite{Wolfenstein:1977ue,Mikheyev:1985zog} occurs when two matter eigenstates become nearly degenerate. We expect two  matter layers for which resonant flavor conversion can occur, each  associated with one of the vacuum frequencies: $\Delta m^2_{21}/2E$ and $\Delta m^2_{31}/2E$.
The  value of $\rho Y_e$ at which resonant flavor conversion takes place is~\cite{Dighe:1999bi}:
\begin{align}
\label{eq:msw_conditions1}
\rho Y_e\big|_{\rm MSW(L)} &= \frac{\Delta m_{21}^2 m_N}{2\sqrt{2}G_FE}\cos2\theta_{12} \simeq 12.87\left(\frac{14 \,{\rm MeV}}{E}\right){\rm \, g \, cm^{-3}}\quad \text{for MSW-L}\, ,\\
\rho Y_e \big|_{\rm MSW(H)}&=\frac{|\Delta m_{31}^2| m_N}{2\sqrt{2}G_FE}\cos2\theta_{13} \simeq 1.15\times10^3\left(\frac{14 \,{\rm MeV}}{E}\right){\rm \, g \, cm^{-3}}  \quad \text{ for MSW-H}   \,  .
\label{eq:msw_conditions2}
\end{align}
Since $|\Delta m^2_{31}| \gg \Delta m^2_{21}$, we refer to these resonances  as MSW(H) and MSW(L), where H and L indicate that the resonances occur at high and low baryon density, respectively. 

\begin{figure}
  \centering
  \includegraphics[width=\linewidth]{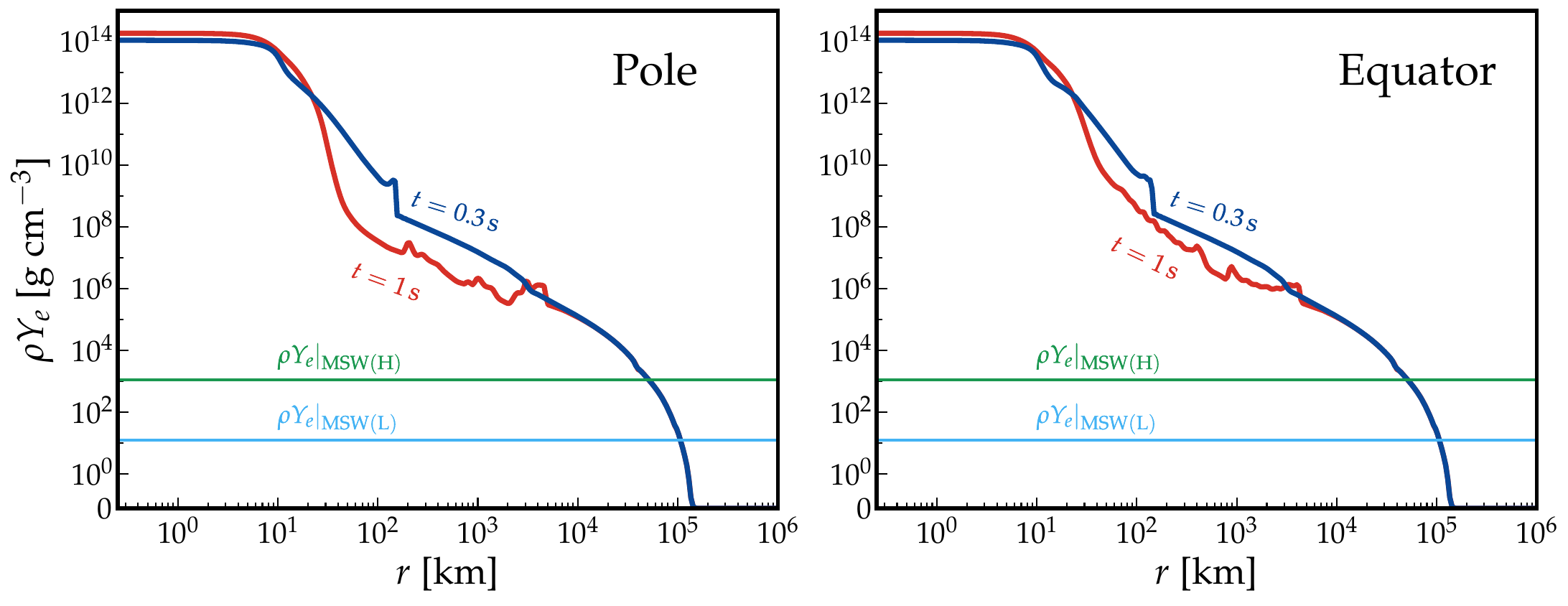}
\caption{Radial profiles of $\rho Y_e$ for our magnetorotational collapse model at $t=0.3$~s (blue) and $t=1$~s (red), shown along the pole ($\theta=0$ and $\varphi = 0$, left panel) and the equatorial direction ($\theta=\pi/2$ and $\varphi = 0$, right panel). 
The horizontal lines indicate the MSW(H)  and MSW(L) resonance conditions (cf.~Eqs.~\ref{eq:msw_conditions1} and \ref{eq:msw_conditions2}) in green and light blue, respectively. 
The resonance conditions have been computed  using a representative neutrino energy corresponding to the time-averaged mean energy of all species in each direction: $E \simeq 14.4$~MeV for the polar direction and $E = 14.2$~MeV for the equator (see Fig.~\ref{fig:lum_en}). 
The radii where the MSW resonances are expected are determined by the intersection between  $\rho Y_e$ and the MSW(H)  and MSW(L) resonance conditions. 
The resonance conditions are not significantly different for the selected post-bounce times and emission directions. The resonances occur in the proximity of  $r_{\mathrm{MSW(H)}} \simeq 5\times10^{4}$~km and $r_{\mathrm{MSW(L)}} \sim 10^{5}$~km. 
} \label{fig:resonance_positions_MSW}
\end{figure}

The effective matter potential has a different  sign for neutrinos and antineutrinos. Hence,  the MSW resonance condition may be satisfied either in the neutrino or the antineutrino sectors, depending on the sign of the relevant mass splitting and   mass ordering.   In our case, the MSW(L) resonance associated with $\Delta m^2_{21}$ always occurs in the neutrino channel, enhancing $\nu_e \leftrightarrow \nu_\mu'$~\footnote{Here, the effective flavor eigenstates ($\nu_\mu'$, $\nu_\tau'$ as well as $\bar{\nu}_\mu'$ and $\bar{\nu}_\tau'$)  diagonalize the $\mu$--$\tau$ sector of the effective Hamiltonian in vacuum.} conversion for both NO and IO.
Conversely, the MSW(H) resonance depends on the sign of  $\Delta m^2_{31}$. It takes place for   neutrinos, if $\Delta m^2_{31}>0$ (NO), leading to $\nu_e \leftrightarrow \nu_\tau'$ conversion; it occurs  for antineutrinos, if $\Delta m^2_{31}<0$ (IO), leading to $\bar{\nu}_e \leftrightarrow \bar{\nu}_\tau'$. 

Figure~\ref{fig:resonance_positions_MSW} displays  the radial profile of $\rho Y_e$ along the polar and  equatorial directions of our magnetorotational collapse model for our two  time snapshots. The regions where the  MSW resonances are expected to take place are approximately marked by the intersection of $\sqrt{2}G_F\,\rho Y_e/m_N$ with $\Delta m^2_{21, 31}/2E$ marked by MSW(L, H), respectively. Resonant flavor conversion is expected to occur  around $10^4$--$10^5$~km. At our selected post-bounce times, the outer stellar layers have not been influenced by the core collapse, therefore there is almost no  difference between the resonance regions along the pole and the equator, and  the different time snapshots.

\begin{figure}[t]
  \centering
\includegraphics[width=\linewidth]{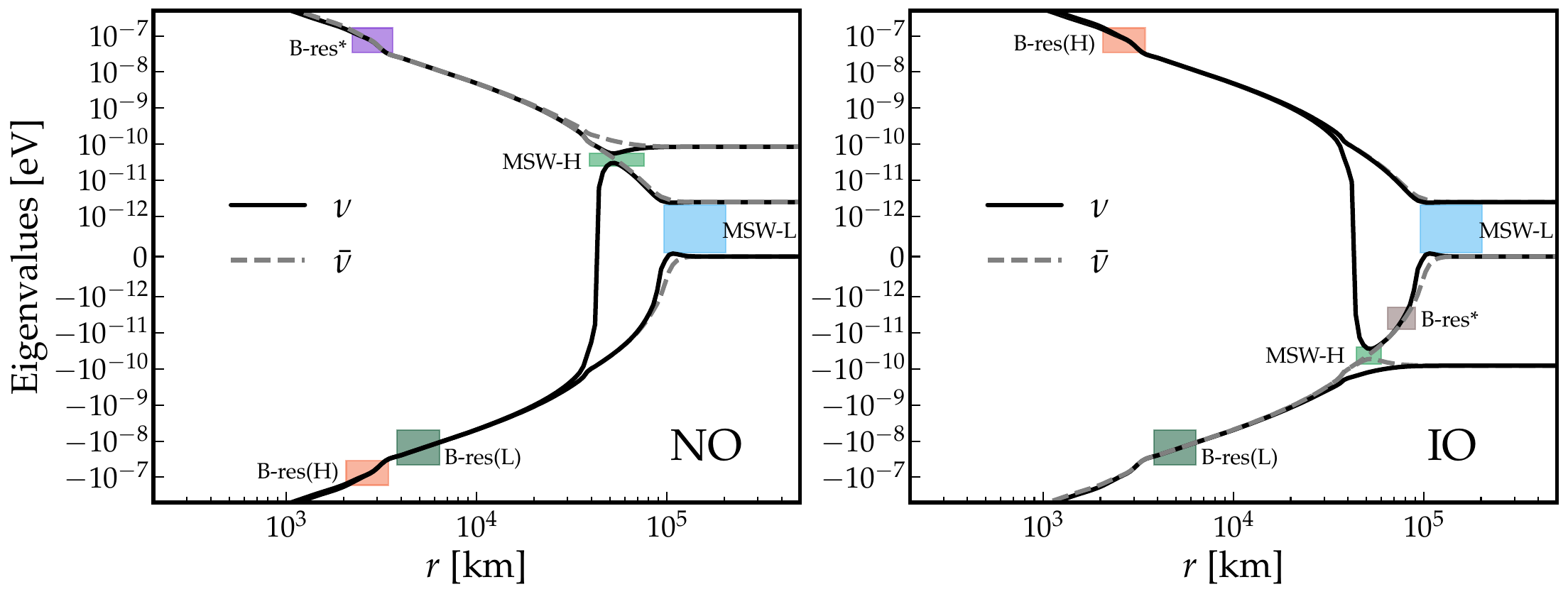}
 \caption{Level-crossing diagrams for NO  (left) and IO  (right). The plots are computed for the equatorial direction ($\theta=\pi/2,\, \varphi = 0$) at $t = 0.3\,\text{s}$ using a neutrino energy $E = 14.2\,\text{MeV}$, which represents the time averaged mean energy across all species (see Fig.~\ref{fig:lum_en}). The features of the level-crossing diagrams are qualitatively invariant across different propagation directions, time snapshots, or energies within the relevant range. For the calculation of the eigenstates relevant for the B-res resonances, a magnetic coupling of $\mu  = 10^{-13} \mu_B$ has been assumed. 
 For $Y_e<0.5$, the MSW(L) resonance occurs in the neutrino channel for both  NO and IO. On the other hand, the MSW(H) resonance takes place in the neutrino (antineutrino) channel for NO (IO), as highlighted by the corresponding level crossings.}
  \label{fig:level_crossing}
\end{figure}

Figure~\ref{fig:level_crossing} represents the level-crossing diagram for NO and IO, i.e., the evolution of the eigenvalues of the Hamiltonian (cf.~Eq.~\ref{eq:hamiltonian3}). The regions where the MSW(L) and MSW(H) resonances occur are marked by green and blue  boxes.
Table~\ref{tab:recap_resonances} summarizes the different types of resonant flavor conversion taking place in our magnetorotational  model and the typical radius at which they occur.

The efficiency of resonant MSW conversion is determined by the adiabaticity parameter~\cite{Landau:1932,Zener:1932}:
\begin{align}
    \gamma_{\rm MSW}= 
\frac{1}{2E}\,
\begin{cases}
    \displaystyle\Delta m^2_{21} \frac{\sin^2 2\theta_{12}}{\cos 2\theta_{12}}\left|\frac{d\ln (\rho Y_e)}{dr}\right|_{r=r_{\rm MSW(L)}}^{-1}  \quad &\text{for MSW(L)\, ,}\\[3ex]
    \displaystyle|\Delta m^2_{31}|\frac{\sin^2 2\theta_{13}}{\cos 2\theta_{13}} \left|\frac{d\ln (\rho Y_e)}{dr}\right|_{r=r_{\rm MSW(H)}}^{-1} \quad &\text{for MSW(H)\, .}
\end{cases}
\end{align}
\begin{table}
\renewcommand{\arraystretch}{1.4}
\caption{Summary of the resonant conversions expected for NO and IO for Majorana particles.
The listed B-res correspond to $Y_e<0.5$; for $Y_e>0.5$, similar resonances occur swapping $\nu \leftrightarrow \bar{\nu}$.
The typical resonance radii are approximately the same for the selected post-bounce times and directions. In contrast, the magnetic field varies more significantly, as shown in Fig.~\ref{fig:profiles}. For this reason, we report the magnetic field values  for the pole and the equator, and provide a variability range to account for the time interval considered.}
\centering
\begin{tabular}{l c c c c c}
\hline\hline
& \multicolumn{2}{c}{Transition} & Typical radius [km] & \multicolumn{2}{c}{Magnetic field $B_\perp$ [G]} \\
& NO & IO & & Pole & Equator \\
\hline
MSW(L) 
 & $\nu_e \leftrightarrow \nu'_\mu$
 & $\nu_e \leftrightarrow \nu'_\mu$
 & $10^5$
 & \multicolumn{2}{c}{Not relevant} \\

MSW(H) 
 & $\nu_e \leftrightarrow \nu'_\tau$
 & $\bar{\nu}_e \leftrightarrow \bar{\nu}'_\tau$
 & $5 \times10^4$ 
 & \multicolumn{2}{c}{Not relevant} \\

\hline
B-res(L) 
 & $\bar{\nu}_e \leftrightarrow \nu'_\mu$
 & $\bar{\nu}_e \leftrightarrow \nu'_\mu$
 & $5 \times10^3$
 & ($1\!-\!3$) $\times10^{10}$
 & $4\times10^{9}\!-\!1\times10^{10}$ \\

B-res(H) 
 & $\bar{\nu}_e \leftrightarrow \nu'_\tau$
 & $\nu_e \leftrightarrow \bar{\nu}'_\tau$
 & $3 \times10^3$
 & ($2\!-\!3$) $\times10^{11}$
 & $1\times10^{10}\!-\!2\times10^{11}$ \\

\multirow{2}*{B-res$^*$}
 & $\nu_e \leftrightarrow \bar{\nu}'_\mu$ & &
 $3 \times10^3$
 & ($2\!-\!3$) $\times10^{11}$
 & $1\times10^{10}\!-\!2\times10^{11}$ \\

 & & $\bar{\nu}_e \leftrightarrow \nu'_\mu$
 & $9 \times10^4$
 & $6\times10^{8}$
 & $6\times10^{8}$ \\
\hline\hline
\end{tabular}
\label{tab:recap_resonances}
\end{table}
Large values of the adiabaticity parameter ($\gamma_{\rm MSW} \gg 1$) indicate  adiabatic flavor evolution, such that the neutrino state follows the instantaneous matter eigenstate across the resonance. Conversely, $\gamma_{\rm MSW} \lesssim 1$ signals non–adiabatic or partially non–adiabatic conversion~\cite{Kuo:1988pn}. We find that the  MSW(L) and MSW(H) resonances are adiabatic in the energy range of interest ($1$--$100$~MeV, along all emission directions and at all post-bounce times for our benchmark magnetorotational model).

In Eqs.~\ref{eq:hamiltonian31} and \ref{eq:hamiltonian3}, we also account for $V_{\mu\tau}$; this term breaks the degeneracy between the $\mu$ and $\tau$ flavors and can trigger an MSW resonance~\cite{Esteban-Pretel:2007ncu}. We find that these corrections do not significantly alter the primary MSW resonances. Furthermore, assuming that the energy spectra of  all non-electron species  are identical (see  Fig.~\ref{fig:lum_en}), any $\nu_\mu \leftrightarrow \nu_\tau$ conversion leaves the final fluxes unchanged, making these corrections negligible.

\subsection{Neutrino-antineutrino   conversion}
\label{sec:Bres}
Since at least two neutrinos are massive, they are also supposed to have a non-zero magnetic moment (see Ref.~\cite{Giunti:2024gec} for a recent review). Hence, in the presence of external magnetic fields, (anti)neutrinos experience dipolar magnetic interactions, which flip their chirality (weak interactions are chiral, meaning that we find only left-handed neutrinos and right-handed antineutrinos).

For Majorana particles, a chirality flip turns neutrinos into antineutrinos and vice versa. Hence, the interaction of the neutrino transition (flavor off-diagonal) magnetic moments with an external magnetic field can  induce a simultaneous change of flavor and chirality, $\nu_\alpha \leftrightarrow \bar{\nu}_\beta$ with 
$\alpha\neq\beta$~\cite{Giunti:2014ixa, Akhmedov:1988uk, Akhmedov:2003fu}. 
The evolution of the system of neutrinos and antineutrinos is therefore described by Eq.~\ref{eq:evolution_eq},  but now the Hamiltonian introduced in Eq.~\ref{eq:HMSW} is modified as follows~\cite{Lim:1988tk,Akhmedov:1989df,Akhmedov:1993sh}:
\begin{align}
    \mathcal{H} = \begin{pmatrix}
        H & \mathcal{B}\\
        \mathcal{B}^\dagger & \bar{H}
    \end{pmatrix}\, .
\end{align}
The effective potential of neutrino interactions with external magnetic fields is 
\begin{align}
    \mathcal{B} =\begin{pmatrix}
        0 & \mu_{e\mu} & \mu_{e\tau} \\
        -\mu_{e\mu} & 0 & \mu_{\mu\tau}\\
        -\mu_{e\tau} & -\mu_{\mu\tau} & 0
    \end{pmatrix} B_\perp\, ,
\end{align}
where $B_\perp$ is the magnetic field component transverse to the neutrino propagation and $\mu_{\alpha\beta}$ are the neutrino magnetic moments.

Flavor conversion is enhanced when a pair (or more) of effective eigenstates becomes close to degenerate. Due to the different sign in the matter potential between neutrinos and antineutrinos, these flavor-changing chirality-flipping conversions can be resonant~\cite{Akhmedov:1988uk,Lim:1988tk}. This phenomenon is  referred to as  resonant spin-flavor precession~\cite{Lim:1988tk, Akhmedov:1988uk, Akhmedov:2003fu}. Hereafter, we dub these transitions  {\it magnetically-driven resonant conversion (B-res)} to avoid   confusion, since  MSW flavor conversion can be  interpreted as  resonant precession of a spin vector~\cite{Kim:1987bv}. For simplicity, we assume  equal magnetic moments for all flavors, i.e., $\mu_{e\mu} = \mu_{e\tau} = \mu_{\mu\tau} = \mu$. Here, $\mu_{\alpha\beta}$ denote the transition magnetic moments between neutrino flavors $\alpha$ and $\beta$. The ratio of magnetic moments $\mu_{e\mu}:\mu_{e\tau}:\mu_{\mu\tau}$  depends on the neutrino mass mechanism and underlying symmetries, see e.g.~Refs.~\cite{Voloshin:1987qy,Barbieri:1988nh,Feruglio:2008ht}. Accounting for  differences among the magnetic moments (albeit plagued by large uncertainties to date) would  only affect the adiabaticity of the resonances, but not their location.

In analogy to the MSW effect, we identify two  resonances~\cite{Akhmedov:2003fu}:
\begin{align}
\label{eq:B-res_conditions1}
\rho(1-2Y_e)\Big|_{\text{B-res(L)}} &=  
\frac{\Delta m_{21}^2 \cos 2\theta_{12}}
{2\sqrt{2}\,G_F\,E}\,m_N = 12.87 \left(\frac{14\, {\rm MeV}}{E}\right){\rm g \, cm ^{-3}}\quad &\text{for B-res(L)}\, ,\\
\rho(1-2Y_e)\Big|_{\text{B-res(H)}} &=
\frac{|\Delta m_{31}^2| \cos 2\theta_{13}}
{2\sqrt{2}\,G_F\,E}\,m_N = 1.15\times10^3\left(\frac{14\, {\rm MeV}}{E}\right){\rm g \, cm ^{-3}}\quad &\text{for B-res(H)}\, .
\label{eq:B-res_conditions2}
\end{align}
Notice that the numerical pre-factors are the same as in Eqs.~\ref{eq:msw_conditions1} and \ref{eq:msw_conditions2}.
In neutron-rich environments ($Y_e < 0.5$),  B-res(L) is responsible for $\bar{\nu}_e \leftrightarrow \nu_{\mu}'$ conversions.  B-res(H) is associated with $\bar{\nu}_e \leftrightarrow \nu_{\tau}'$  in NO and $\nu_e \leftrightarrow \bar{\nu}_{\tau}'$ in IO. On the other hand, for proton rich media, neutrinos and antineutrinos are swapped.

\begin{figure}[t]
  \centering
  \includegraphics[width=\linewidth]{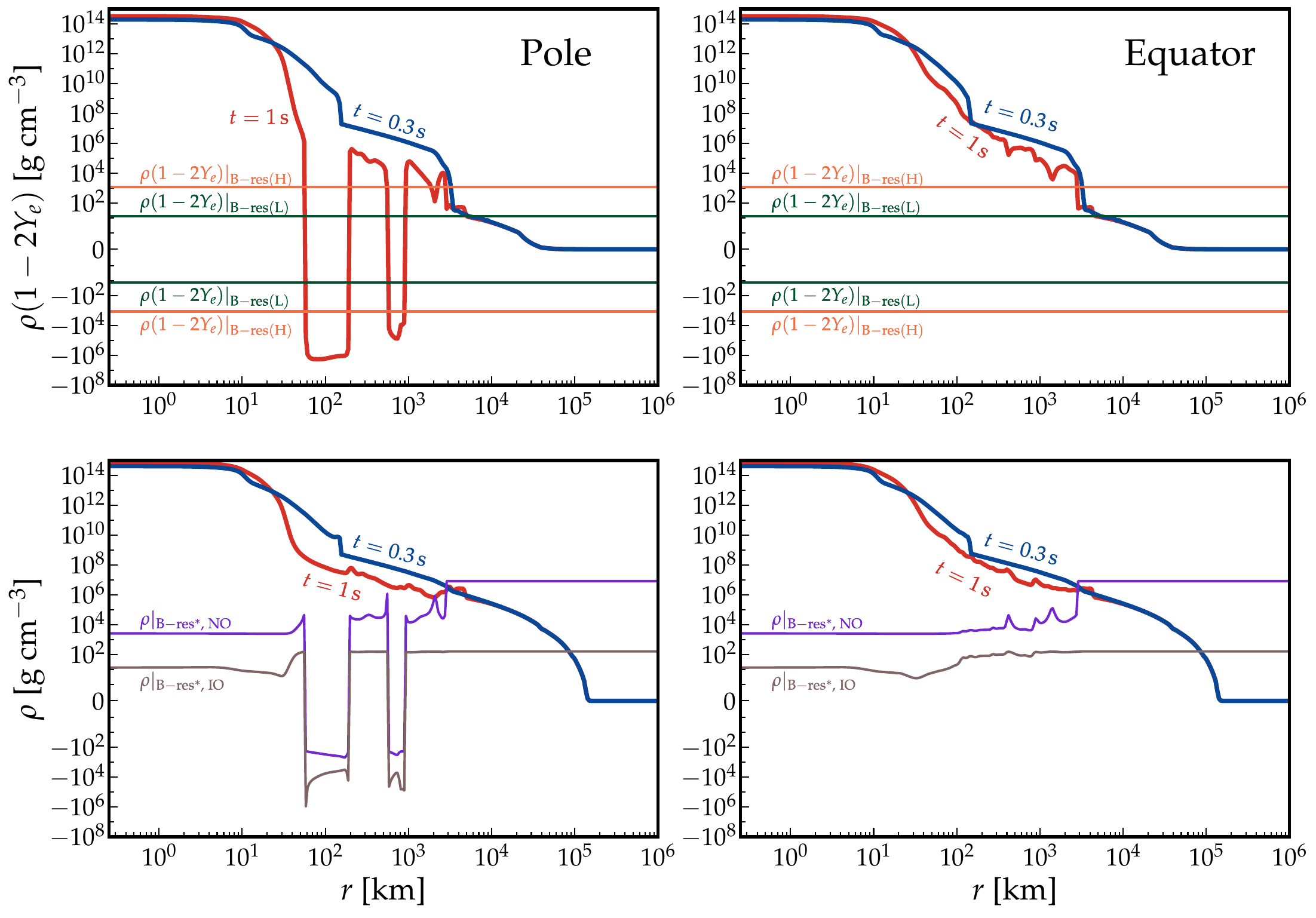}
  \caption{Radial profiles of $\rho (1 - 2 Y_e)$ (top panels) and $\rho$ (bottom panels) for our magnetorotational model at $t = 0.3$~s (blue) and $t = 1$~s (red), along  the polar ($\theta=0,\, \varphi = 0$, left) and equatorial ($\theta=\pi/2,\, \varphi = 0$, right) directions. In the top panels, the horizontal lines mark the  B-res(H) and B-res(L) conditions (see Eqs.~\ref{eq:B-res_conditions1} and \ref{eq:B-res_conditions2}). Negative values of $\rho(1 - 2Y_e)$  indicate regions with $Y_e > 0.5$. In the bottom panels, the purple and gray lines mark the B-res$^*$ condition (Eq.~\ref{eq:B-resstar}). 
The resonance conditions  are computed for  a representative neutrino energy equal to the time-averaged mean energy of all species: $E \simeq 14.4$~MeV for the polar direction and $E = 14.2$~MeV for the equatorial direction (see Fig.~\ref{fig:lum_en}).
}
  \label{fig:resonance_positions_Bres_total}
\end{figure}

Figure~\ref{fig:resonance_positions_Bres_total} (top row) shows the B-res conditions as  functions of radius for the polar (left) and equatorial (right) directions; see also Table~\ref{tab:recap_resonances} that summarizes the transitions occurring at each resonance for NO and IO depending on whether $Y_e < 0.5$ [$\rho(1-2Y_e)>0$] or $Y_e > 0.5$ [$\rho(1-2Y_e)< 0$].
The conditions for  B-res(H) and  B-res(L)  are met at several radii at $t=1$~s in the polar direction.
However, most of these B-res  occur in regions where the matter potential varies extremely rapidly. Such sharp gradients are expected to lead to highly non-adiabatic conditions that disfavor efficient  flavor conversion. Consequently, the physically relevant resonances are confined to a relatively narrow spatial window of a few thousand kilometers. We also note that the B-res conditions, and their adiabaticity, are negligibly affected by the specific  interpolation method employed for magnetic field profile (cf.~Fig.~\ref{fig:profiles}).

In the presence of multiple B-res radii close to each other (where $\rho(1-2Y_e)$ does not vary rapidly),  one should compute the adiabatic parameter at all radii where the B-res conditions are met. However, for simplicity, we only compute the adiabaticity parameter at the largest radius.
In fact, since  the magnetic field strength decreases with distance (see Fig.~\ref{fig:profiles}),  the outermost crossing provides a more conservative estimate for the adiabaticity of the transition. 
Comparing the B-res locations with the $Y_e$ profiles in Fig.~\ref{fig:profiles}, it is evident that these resonances occur for $Y_e \simeq 0.5$.

Another resonance can take place that we denote as B-res$^*$ (referred to as RSFP-E in Ref.~\cite{Akhmedov:2003fu}). 
Specifically, it leads to the $\nu_e \leftrightarrow \bar{\nu}_{\mu}'$ conversion in NO and the $\bar{\nu}_e \leftrightarrow \nu_{\mu}'$  in IO. 
A convenient analytic estimate of the baryon density at which  B-res$^*$ occurs can be obtained by expanding the three flavor eigenvalues as  functions of  $(1-2Y_e)$ and $\Delta m^2_{21}/\Delta m^2_{31}$~\cite{Akhmedov:2003fu}:
\begin{align}
    \rho\big|_{\rm B-res^*} = \frac{|\Delta m_{31}^2| \Bigl[(1 \mp a + z)\pm \sqrt{(1 \mp a + z)^2 \pm 4a}\Bigr]} 
{2\sqrt{2}\,G_F\,E}\,m_N \, , 
\label{eq:B-resstar}
\end{align}
where the upper (lower) signs refer to NO (IO). The parameters $a$ and $z$ are defined as
\begin{equation}
a \equiv \frac{Y_e}{1-2Y_e}\,\frac{\Delta m_{21}^2}{|\Delta m_{31}^2|}\,\cos 2\theta_{12}\, ,
\qquad
z \equiv \sin\theta_{13}\,\frac{Y_e}{1-2Y_e}\, .
\label{eq:az_def}
\end{equation}
Equation~\eqref{eq:B-resstar} is an approximate expression whose precision depends  on whether $1-2Y_e \ll 1$~\cite{Akhmedov:2003fu}. 
Notice that the variables $a$ and $z$  depend on  radius; therefore, it is not possible to give a radius-independent value of the resonance conditions, contrary to the other resonances. 

The bottom row of Fig.~\ref{fig:resonance_positions_Bres_total} shows Eq.~\eqref{eq:B-resstar}. We can see that  B-res$^*$  occurs in the same spatial region as  B-res(H/L) for NO (cf.~top row of Fig.~\ref{fig:resonance_positions_Bres_total}). Conversely,   B-res$^*$ shifts to larger radii for IO, overlapping with the MSW resonances. 
Figure~\ref{fig:level_crossing} summarizes the interplay between MSW resonances, B-res(H), B-res(L), and B-res$^*$; the colored boxes mark the resonance locations.
Note, however, that the  B-res level crossings are not sharply visible in the eigenvalue diagram~\footnote{In Refs.~\cite{Akhmedov:2003fu, Jana:2022tsa}, the B-res crossings are plotted in arbitrary units; this choice  effectively magnifies the local crossings gaps. Here, instead, we plot the eigenvalues in physical units, hence  the  small B-res gap.}. 

We assess the efficiency of the B-res transitions by means of  the adiabaticity parameter~\cite{Akhmedov:2003fu}:
\begin{align}
    \gamma_{\rm B-res} = 8E(\mu B_\perp)^{2} \begin{cases}
        \displaystyle \frac{1}{\Delta m^2_{21}}\left|\frac{1}{\rho(1-2Y_e)} \frac{d\bigl[\rho\,(1-2Y_e)\bigr]}{dr} \right|^{-1}_{r=r_{\rm B-res(L)}} &\text{for B-res(L)}\, ,\\[5pt]
\displaystyle \frac{1}{\left|\Delta m^2_{31}\right|}\left|
\frac{1}{\rho(1-2Y_e)}
\frac{d\bigl[\rho\,(1-2Y_e)\bigr]}{dr}
\right|^{-1}_{r=r_{\rm B-res(H)}} &\text{for B-res(H)}\, ,\\[5pt]
\displaystyle \frac{1}{\sin_{13}^2|\Delta m^2_{31}|-\cos 2\theta_{12}\Delta m_{21}^2}\left|\frac{1}{\rho(1-Y_e)} \frac{d\bigl[\rho\,(1-Y_e)\bigr]}{dr} \right|^{-1}_{r=r_{\rm B-res ^*}} &\text{for B-res$^*$}\, .
    \end{cases}
    \label{eq:gamma_bres}
\end{align}
The adiabaticity parameter depends on  $\mu\,B_\perp$ (the magnetic field perpendicular to the neutrino propagation)  at the resonance radius. This is because, since we focus on spatial regions away from the neutrinospheres,  we assume that  neutrinos  propagate freely in the radial direction~\cite{Tamborra:2017ubu}; therefore, the relevant magnetic field component is the one perpendicular to the radial direction, $B_\perp$ 
(cf.~Sec~\ref{sec:mrcc}). For each resonant channel, if $\gamma_{\rm{B-res}} \gg 1$,  flavor conversion at the corresponding resonance is adiabatic, whereas  the transition is non-adiabatic for $\gamma_{\rm{B-res}} \ll 1$ (i.e., chirality-flipping flavor conversion is not efficient).

\begin{figure}[t]
  \centering
\includegraphics[width=\linewidth]{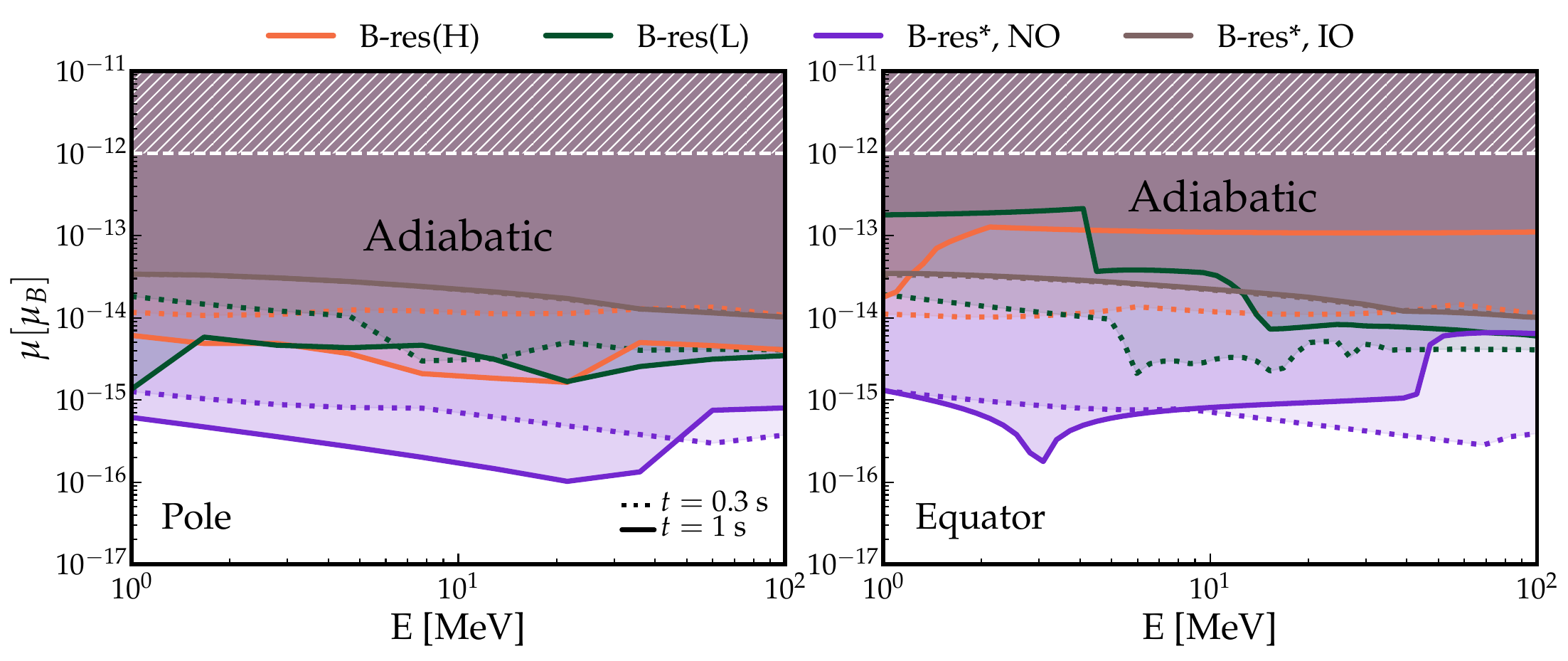}
\caption{Values of the neutrino magnetic moment $\mu$ such that the adiabaticity parameter $\gamma \simeq 1$ (Eq.~\ref{eq:gamma_bres})  for the polar ($\theta=0$ and $\varphi = 0$, left) and equatorial ($\theta=\pi/2$ and $\varphi = 0$, right) directions. The solid (dotted) curves correspond to $t=1$~s ($t=0.3$~s). For each resonance, the shaded areas (above the corresponding curve)  satisfy $\gamma \gtrsim 1$, hence  adiabatic transitions should be expected. 
The resonance conditions  are computed for  a representative neutrino energy corresponding to the time-averaged mean energy of all species: $E \simeq 14.4$~MeV for the polar direction and $E = 14.2$~MeV for the equatorial direction (see Fig.~\ref{fig:lum_en}). 
The resonances are adiabatic even for values of $\mu$ well below the most stringent experimental bounds, i.e., $\mu \lesssim  10^{-12}\mu_B$~\cite{Giunti:2014ixa,ParticleDataGroup:2024cfk}, shown as the white hatched region.} 
  \label{fig:mu_limits} 
\end{figure}

Figure~\ref{fig:mu_limits}  shows the range of the neutrino magnetic moment such that  $\gamma_{\rm{B-res}} \gtrsim 1$ for all types of B-res (shadowed regions, see Eq.~\ref{eq:gamma_bres}), at $t=0.3$~s and $t=1$~s and along the polar and equatorial directions. 
The diverse behavior along the polar and equatorial directions, and for the different B-res types, stems from the profile of $B_\perp$ (cf.~Fig.~\ref{fig:profiles}).  The resonances are adiabatic even for values of $\mu$ smaller than the most stringent experimental bounds, $\mu \lesssim 10^{-12} \mu_B$~\cite{Giunti:2014ixa,ParticleDataGroup:2024cfk}. The typical values of  $B_\perp$ at the resonance locations  are summarized in Table~\ref{tab:recap_resonances}. 
We have also solved the equations of motion numerically and cross checked that the analytical expressions provided in this section are in overall excellent agreement with the numerical solution.

\subsection{Discussion}
We investigate the role of MSW and B-res flavor transitions in a magnetorotational core collapse, relying  on a three-dimensional neutrino-magnetohydrodynamic simulation (see Sec.~\ref{sec:mrcc} for details) and therefore employing a  realistic modeling  of the magnetic field. This approach reveals that the strong magnetic field ($B_\perp \simeq 10^{15}$--$10^{16}$~G) is responsible for making all B-res transitions adiabatic for  neutrino magnetic moments compatible with the experimental limits~\cite{Giunti:2014ixa,ParticleDataGroup:2024cfk}. Moreover, the strong directional dependence characteristic of the three-dimensional magnetorotational collapses  further impacts  flavor evolution. In fact, all quantities relevant for exploring flavor conversion depend on the emission angles $(\theta, \varphi)$: both the MSW and B-res features, and their adiabaticity, change with the emission direction; this translates into modified fluxes at Earth, as shown in Secs.~\ref{sec:flux_conversion} and \ref{sec:detection}.

Existing literature on the topic employs simplified dipolar configurations of the magnetic field with analytical radial dependence~\cite{Ando:2002sk,Ando:2003is,Ando:2003pj,Akhmedov:2003fu, Jana:2022tsa, deGouvea:2012hg}. 
Hence, our findings present a key difference with respect to previous work, see e.g.~Refs.~\cite{Ando:2002sk,Ando:2003is,Jana:2022tsa}, where the B-res$^*$ and B-res(L) transitions were found to be non-adiabatic due to the weaker magnetic field profiles, characteristic of neutrino-driven core-collapse supernovae. The corresponding resonances typically occur at radii separated by almost an order of magnitude in Refs.~\cite{Ando:2002sk,Ando:2003is,Jana:2022tsa}, whereas they are much closer to each other in our case, owing to the different $\rho(1-2Y_e)$ profile (see Fig.~\ref{fig:resonance_positions_Bres_total} and Table~\ref{tab:recap_resonances}).

We focus on exploring the flavor conversion phenomenology for Majorana particles. 
However, for Dirac neutrinos, the chirality flip resulting from dipolar magnetic interactions implies left-handed neutrinos turn into right-handed ones (and right-handed antineutrinos into left-handed ones), which then evade detection via weak interactions. For Dirac neutrinos, matter can also enhance these conversions resonantly. The resonance location would be different with respect to the one investigated for Majorana neutrinos.  As a result, neutrino magnetic dipole interactions may be responsible for  a reduction of the detected flux for Dirac particles. This scenario may be degenerate with other uncertainties plaguing the expected flux at Earth, such as the core-collapse physics, the role of neutrino-neutrino interaction, and potential beyond-the-Standard-Model scenarios. For more details, we refer the reader to Ref.~\cite{Giunti:2014ixa}, where the phenomenology of B-res transitions is  thoroughly addressed. Related resonant flavor conversion mechanisms for Dirac neutrinos induced by twisting magnetic fields have also been explored in Ref.~\cite{Jana:2023ufy}.

\section{Flavor composition at Earth} \label{sec:flux_conversion}
The neutrino flux seen by an observer on Earth is the result of weighted hemispheric averages that take into account projection effects in the observer direction of the oscillated, direction- and flavor-dependent fluxes emitted at the source. Hence,  the flavor-dependent neutrino fluxes observed at Earth along the polar or equatorial direction (in the same coordinate system of the neutrino-magnetohydrodynamic simulation) are obtained computing the oscillated fluxes emitted under different angles at the source, and then projecting them along the observer direction (as outlined in Appendix A of Ref.~\cite{Tamborra:2014hga}), and employing   Eq.~\eqref{eq:flux} for each emission direction at the source.

As discussed in Sec.~\ref{sec:MSW}, the MSW resonances are  adiabatic at all post-bounce times and along all emission directions for our magnetorotational core-collapse model. Therefore, the only potential source of non-adiabaticity arises from B-res, whose behavior is parametrized relying on the Landau-Zener jump probability~\cite{Landau:1932,Zener:1932}:
\begin{equation}
p_i \simeq \exp\!\left[-\frac{\pi}{2}\,\gamma_{{\rm B-res},i}\right],
\qquad i=\rm{L}, \rm{H}, *\, ,
\label{eq:pi_definition}
\end{equation}
with the adiabaticity parameter, $\gamma_{{\rm B-res},i}$, defined as in Eq.~\eqref{eq:gamma_bres}. 
Here, $p_i=0$ corresponds to a fully adiabatic crossing and $p_i=1$ to a  non-adiabatic crossing. Figure~\ref{fig:mu_limits} shows that, for magnetic moments below the  experimental bounds ($10^{-13}\mu_B \lesssim \mu \lesssim 10^{-12}\mu_B$), $\gamma_{\rm B-res} \gtrsim 1$ along  the polar and equatorial directions. Hence, the B-res transitions are adiabatic. 
Conversely, if magnetic moments should be  extremely small or vanishing (i.e., $\mu \lesssim 10^{-16} \mu_B$),  B-res transitions are  non-adiabatic;  only the standard MSW resonant conversion affects the flavor ratio.

The (anti)neutrino fluxes expected at Earth in NO, assuming
adiabatic MSW transitions  (Sec.~\ref{sec:MSW}), accounting for the loss of coherence when travelling to Earth, and
allowing for generic values of the B-res jump probabilities $p_i$,
are:
\begin{align}
\Phi_{\nu_e}=\,&
|U_{e3}|^2 p_{*}\,\Phi^0_{\nu_e}
+\left[ |U_{e1}|^2 (1-p_{L})p_{H} + |U_{e2}|^2 (1-p_{H})\right]\Phi^0_{\bar{\nu}_e} \nonumber\\
&+\left[ |U_{e1}|^2\!\left((1-p_L)(1-p_H)+p_L\right)
+ |U_{e2}|^2 p_H
+ |U_{e3}|^2 (1-p_{*}) \right]\Phi^0_{\nu_x}\, ,
\label{eq:phinue_NO_general}
\\[2mm]
\Phi_{\bar{\nu}_e}=\,&
|U_{e2}|^2(1-p_{*})\,\Phi^0_{\nu_e}
+|U_{e1}|^2 p_L p_H\,\Phi^0_{\bar{\nu}_e} \nonumber\\
&+\left[ |U_{e1}|^2\!\left((1-p_L)+p_L(1-p_H)\right)
+|U_{e2}|^2 p_{*}
+|U_{e3}|^2 \right]\Phi^0_{\nu_x}\, ,
\label{eq:phinuebar_NO_general}
\end{align}
where  $U_{ei}$ ($i=1,2,3$) represents the  elements of the PMNS mixing matrix    linked to the electron flavors (cf.~Sec.~\ref{sec:MSW}).
For IO, we have
\begin{align}
\Phi_{\nu_e}=\,&
\left[ |U_{e1}|^2 (1-p_{*})(1-p_H) + |U_{e2}|^2 p_H\right]\Phi^0_{\nu_e}
+|U_{e1}|^2 (1-p_L) p_{*}\,\Phi^0_{\bar{\nu}_e} \nonumber\\
&+ \left[ |U_{e1}|^2\!\left((1-p_{*})p_H+p_{*}p_L\right)
+ |U_{e2}|^2 (1-p_H)
+ |U_{e3}|^2 \right]\Phi^0_{\nu_x}\, ,
\label{eq:phinue_IO_general}
\\[2mm]
\Phi_{\bar{\nu}_e}=\,&
|U_{e1}|^2 (1-p_H) p_{*}\,\Phi^0_{\nu_e}
+\left[ |U_{e1}|^2 (1-p_{*})(1-p_L)
+|U_{e3}|^2 p_L \right]\Phi^0_{\bar{\nu}_e} \nonumber\\
&+\left[ |U_{e1}|^2\!\left((1-p_{*})p_L+p_{*}p_H\right)
+|U_{e2}|^2
+|U_{e3}|^2 (1-p_L) \right]\Phi^0_{\nu_x}\, .
\label{eq:phinuebar_IO_general}
\end{align}
The flux of non-electron flavors follows from particle-number conservation:
\begin{equation}
\Phi_{\nu_x}=
\Phi^0_{\nu_e}+\Phi^0_{\bar{\nu}_e}+4\Phi^0_{\nu_x}
-\Phi_{\nu_e}-\Phi_{\bar{\nu}_e}\, .
\label{eq:phix_conservation}
\end{equation}

In what follows, we focus on two scenarios.
\begin{itemize}
    \item \textit{Adiabatic MSW conversion}
    (obtained by setting $p_L = p_H = p_* = 1$).
    This limit holds for small magnetic moments  (i.e., $\mu \lesssim 10^{-16} \mu_B$), such that  all  B-res transitions are  non-adiabatic.
    \item \textit{Adiabatic MSW and  B-res conversion}
    (for $p_L = p_H = p_* = 0$).  This scenario is to be expected in the presence of large  transition magnetic moments for Majorana neutrinos (e.g.~$\mu = 10^{-12} \mu_B$, compatible with experimental constraints); all B-res resonances are  adiabatic.
\end{itemize}

\begin{figure}[t]
  \centering
\includegraphics[width=\linewidth]{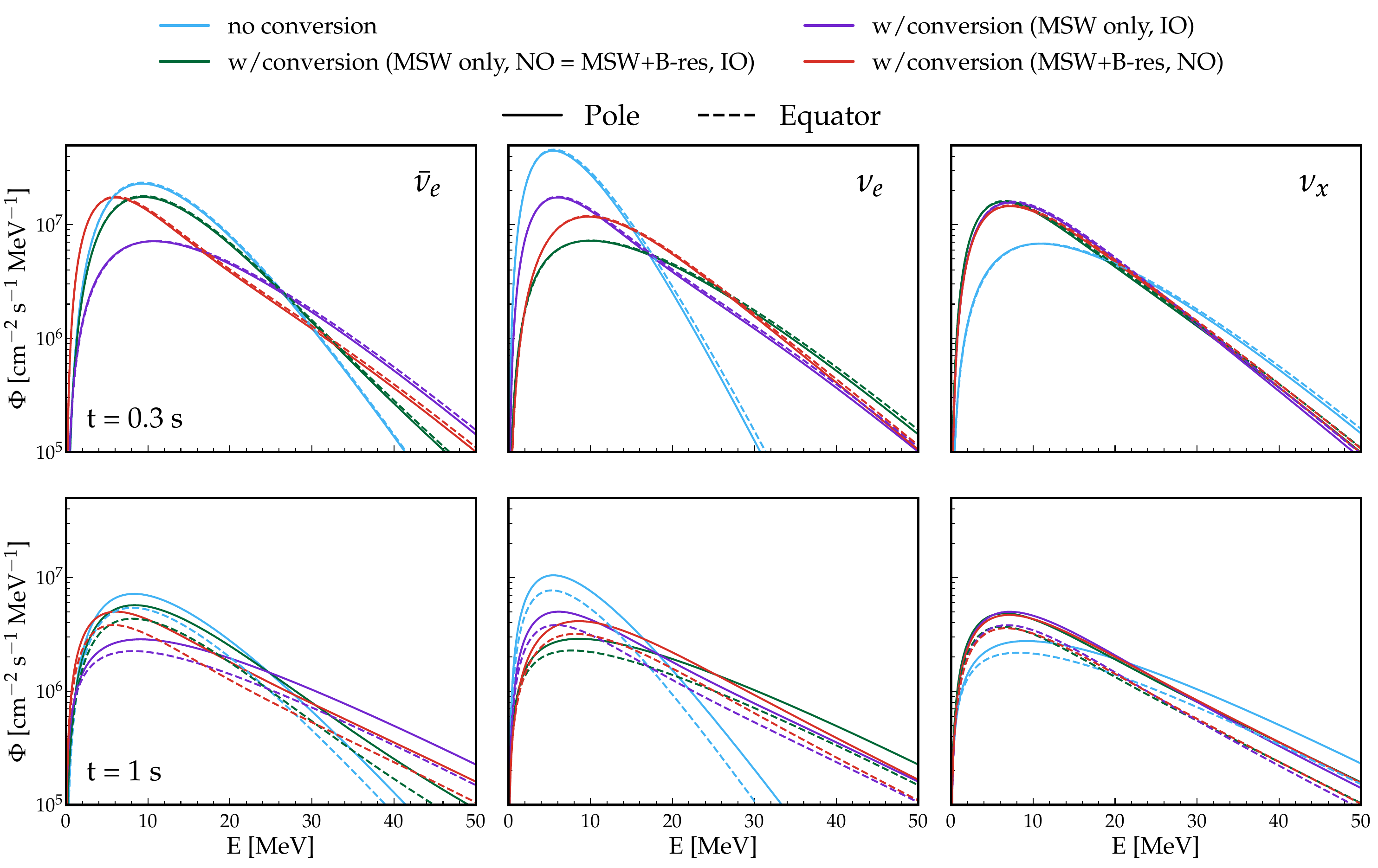}
\caption{Flavor-dependent neutrino fluxes observable at Earth for an observer located along the pole ($\theta=0,\, \varphi = 0$, solid lines) and the equator ($\theta=\pi/2,\, \varphi = 0$, dashed lines) for $t=0.3$~s (top panels) and $t=1$~s (bottom panels). The fluxes of $\bar\nu_e$, $\nu_e$, and $\nu_x$ are plotted from left to right, respectively. The light-blue  curves represent the projected fluxes in the absence of flavor conversion (Eq.~\ref{eq:flux}). The olive (purple) lines show the fluxes in the presence of MSW conversion in NO as from Eqs. \ref{eq:phinue_NO_general}-\ref{eq:phinuebar_NO_general} with $p_i=0$ (IO, Eqs. \ref{eq:phinue_IO_general}-\ref{eq:phinuebar_IO_general} with $p_i=0$); the MSW-NO fluxes are degenerate with the ones  obtained  in the presence of MSW and B-res flavor conversion in IO  (Eq.~\ref{eq:phinue_IO_general} and Eq.~\ref{eq:phinuebar_IO_general} with $p_i=1$). The red curves represent the fluxes  in the presence of MSW and B-res flavor conversion in NO (Eq.~\ref{eq:phinue_NO_general} and Eq.~\ref{eq:phinuebar_NO_general} with $p_i=1$). The oscillated fluxes display a more pronounced dependence on the observer direction for $t=1$~s.}
\label{fig:flux_earth}
\end{figure}
Figure~\ref{fig:flux_earth} shows the flavor-dependent fluxes expected for an observer located along the pole (solid lines) and the equator (dashed lines) at $t=0.3$~s (top panels) and $t=1$~s (bottom panels). The neutrino fluxes depend more prominently on the observer direction for $t=1$~s, whereas a negligible directional dependence is visible for $t=0.3$~s. This is due to the development of narrow jets  along the polar directions over time (cf.~Fig.~9 of Ref.~\cite{Obergaulinger:2021omt}), leading to largely  asymmetric matter ejection. 
Our magnetorotational collapse model does not distinguish between muon and tau (anti)flavors.  Therefore,  some flavor conversion scenarios become  degenerate. For example, $\Phi_{\bar\nu_e}$ with $p_i=1$ in NO coincides with $\Phi_{\bar\nu_e}$ with $p_i=0$ in IO. Importantly, such degeneracy does not imply a fundamental equivalence of the underlying flavor conversion mechanisms, which differ in their resonance structure.

\section{Detection prospects}
\label{sec:detection}
In this section, we compute the event rate  in 
the IceCube Neutrino Observatory~\cite{Abbasi:2011ss} and Hyper-Kamiokande~\cite{Hyper-Kamiokande:2018ofw} expected from a magnetorotational collapse at $D=10$~kpc from Earth. In both detectors, neutrinos are mainly detected through inverse beta decay (IBD, $\bar{\nu}_e+p\rightarrow n+e^+$).  This detection channel accounts for the majority of the total signal, while subleading contributions arise from elastic scattering on electrons as well as charged and neutral current interactions on oxygen~\cite{Abbasi:2011ss}, which we neglect here. For a neutrino-driven core-collapse supernova, these secondary detection channels would contribute less than  $10\%$ to the total rate~\cite{Abbasi:2011ss}; we estimate similar figures for neutrino detection from magnetorotational collapses.

We model the IceCube event rate following Refs.~\cite{Abbasi:2011ss, Tamborra:2014hga}. The IBD-induced rate in a single optical module is
\begin{equation}
    r_{\rm IBD} = n_{\rm p, \,IC} \int dE_e \int dE_\nu \,
\Phi_{\bar\nu_e}(E_\nu)\, \sigma'(E_e,E_\nu)\,
N_\gamma(E_e)\, V_{\rm eff}\, ,
\label{eq:ICE}
\end{equation}
where $n_{\rm p,\, IC} = 6.18 \times 10^{22}\,{\rm cm^{-3}}$ is the number density of free protons in ice, $V_{\rm eff}=0.163 \times 10^{6}\,{\rm cm^3}$ is the effective volume for single photon detection, and $N_\gamma(E_e)=178\,E_e/{\rm MeV}$ is the number of Cherenkov photons produced by a positron of energy $E_e$. The differential cross section $\sigma'(E_e,E_\nu)=d\sigma/dE_e$ is modeled following Ref.~\cite{Strumia:2003zx}. We also shift the positron energy as $E_e\rightarrow E_e+1~{\rm MeV}$ to account  for the additional visible energy from annihilation and neutron-capture photons~\cite{Abbasi:2011ss}. IceCube is made of  $N_{\rm OM}=5160$ optical modules. We incorporate a dead time correction, which accounts for background suppression and the associated reduction of the observable signal, so that the observed signal rate is
\begin{align}
    R_{\rm IC} = N_{\rm OM}\frac{0.87 \,r_{\rm IBD}}{1+ \tau\,r_{\rm IBD}}\, .
\end{align}
Here, $\tau = 250~\mu$s is the dead-time of the optical module, and the factor $0.87$ results from the parametrization of the dead-time effect.

The expected event rate in the water Cherenkov detector Hyper-Kamiokande~\cite{Hyper-Kamiokande:2018ofw} is:
\begin{equation}
    R_{\rm HK} = N_{\rm p,\, HK} \int dE_e \int dE_\nu \,
\Phi_{\bar\nu_e}(E_\nu)\, \sigma'(E_e,E_\nu)\, ,
\label{eq:HK}
\end{equation}
where $N_{\rm p,\,HK} = 1.25 \times 10^{34}$ is the number of free protons, resulting from a simple scaling in size with respect to Super-Kamiokande, as in Ref.~\cite{Martinez-Mirave:2025pnz}.

\begin{figure}[t]
  \centering
\includegraphics[width=\linewidth]{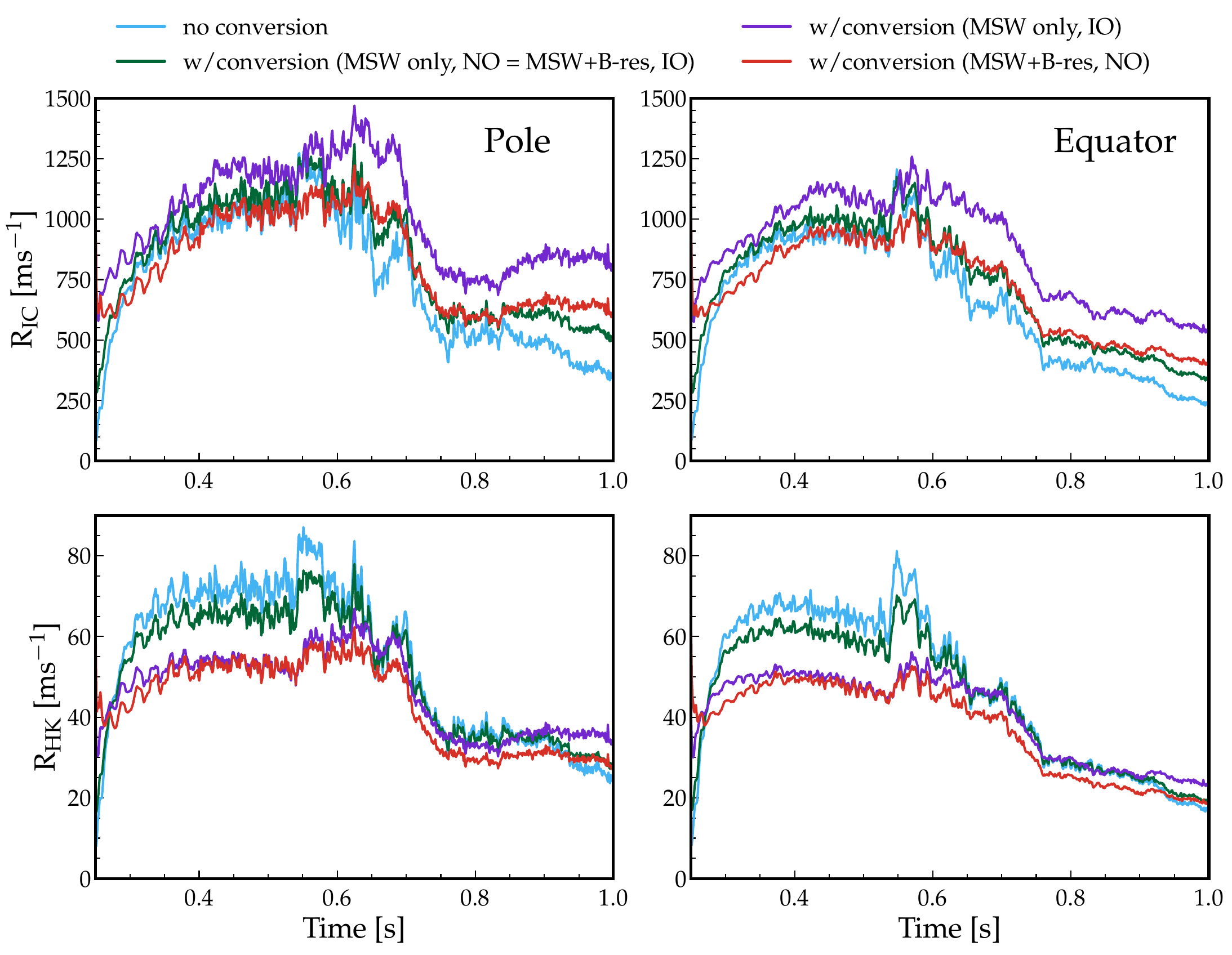}
\caption{IceCube (top panels) and Hyper-Kamiokande (bottom panels) event rates  as  functions of time for a magnetorotational collapse occurring at $D=10$~kpc from Earth. The left and right panels correspond, respectively, to an observer direction close to the  pole ($\theta=0,\, \varphi = 0$) and at the  equator ($\theta=\pi/2,\, \varphi = 0$).
The event rates are computed accounting for the  three-flavor conversion scenarios outlined in Sec.~\ref{sec:flux_conversion}; see Fig.~\ref{fig:flux_earth}. 
The event rates prominently depend on the observer direction, with the rate of events being larger for an observer located at the pole. The different energy dependence of the IceCube and Hyper-Kamiokande rates is responsible for a different  hierarchy among the oscillation scenarios leading to the  largest event rate.  
}
\label{fig:event_rates}
\end{figure}
Figure~\ref{fig:event_rates} shows the detection rate in IceCube (top panels) and Hyper-Kamiokande (bottom panels) for a magnetorotational collapse at $D=10$~kpc from Earth, assuming an observer located at the  pole (left panels) and at the equator (right panels).
In all cases, the signal rapidly rises  after bounce and reaches a broad maximum  at $t \simeq 0.5$--$0.6$~s, followed by a gradual decline as the neutrino luminosity decreases.
The polar direction generally exhibits larger event rates, consistent with the enhanced neutrino emission presented in Sec.~\ref{sec:mrcc}. In particular, IceCube features a significantly larger event rate due to the larger detection volume. Another notable difference between the two detectors arises from their energy dependence. In Hyper-Kamiokande, the event rate is directly proportional to the $\bar\nu_e$ flux and the IBD cross section (Eq.~\ref{eq:HK}), while the IceCube  signal  depends on the number of Cherenkov photons (Eq.~\ref{eq:ICE}).
Flavor conversion  modifies  the normalization and the time evolution of the event rate. 
The MSW (NO) and $B$-res (IO) curves overlap due to the degeneracy of the neutrino fluxes observed in Fig.~\ref{fig:flux_earth}. 
Due to the different energy dependence of the IceCube and Hyper-Kamiokande rates, the hierarchy among the oscillation scenarios leading to the  largest or smallest rate is different.  For instance, the MSW  scenario in IO leads to the largest IceCube rate, since  $\Phi_{\bar{\nu}_e} \sim \Phi^0_{\nu_x}$, which has the largest mean energy (cf. Fig.~\ref{fig:lum_en}); on the other hand, the no flavor conversion case is responsible for the highest event rate in Hyper-Kamiokande. We stress that the event rates may  be further affected by neutrino self-interaction, however we neglect this physics here for simplicity. 

From Fig.~\ref{fig:event_rates}, one can also see that the expected rate exhibits a significant dependence on the observer direction. 
The anisotropy of  antineutrino emission becomes more pronounced as large-scale asymmetries linked to the explosion dynamics develop. However, the directional dependence of the event rate differs slightly in IceCube and Hyper-Kamiokande because of the related energy dependence of the event rate.

\section{Conclusions}
\label{sec:conclusion}
Up to few percent of collapsing massive stars is characterized by fast rotation and large magnetic field. Magnetorotational collapses are expected to emit a large number of neutrinos.
We expect  non-electron flavor (anti)neutrinos from magnetorotational collapses to have average energy much larger than expected in the standard neutrino-driven core-collapse scenario, due to the fact that neutrinos decouple at higher densities.
In this work, we employ a  three-dimensional special relativistic neutrino-magnetohydrodynamical simulation of a $13 M_\odot$ model~\cite{obergaulinger-aloy,Obergaulinger:2021omt} with SFHo nuclear equation of state. For the first time, accounting for the directional dependence of the electron density and magnetic field  profiles at all post-bounce times, we investigate the occurrence of both MSW resonant conversions and neutrino-antineutrino conversion due to magnetic moment transitions in three flavors. The latter form of flavor conversion is expected to occur if neutrinos should be Majorana particles, whereas the former may occur independent of the Dirac or Majorana nature of neutrinos.

We find that resonant MSW flavor conversion  is always adiabatic for any  neutrino emission direction and post-bounce time. 
The large magnetic field present  in the outer layers of the collapsing star is responsible for resonant flavor-changing neutrino-antineutrino conversion. Such transitions are adiabatic for neutrino magnetic moments $\lesssim 10^{-12}\mu_B$ (compatible with  terrestrial and astrophysical constraints), with qualitatively similar resonant behavior for different emission directions of neutrinos.

The (anti)neutrino flux expected at Earth is obtained  accounting for  weighted hemispheric averages of the oscillated (anti)neutrino fluxes emitted from the magnetorotational collapse in the observer direction. We compute the  event rate expected in the IceCube Neutrino Observatory and in Hyper-Kamiokande for an observer located at the pole (i.e.~observing the jet harbored during the magnetorotational collapse head on) and at the equator (i.e.~observing the magnetorotational collapse perpendicularly with respect to the jet direction). We find that the expected event rate strongly depends on the oscillation scenario, and on whether neutrinos are Dirac or Majorana particles. The event rate is generally larger in the polar direction, especially at later post-bounce times.
 
The potential joint detection of neutrinos and gravitational waves from magnetorotational collapses would be crucial to extract key information on the core collapse physics. However, the strong hierarchy in the flavor-dependent neutrino energies calls for a deep  understanding of flavor conversion physics in order to be able to test our theoretical understanding. Our work is a first step in this direction, it highlights a rich phenomenology associated to (anti)neutrinos that can largely impact multi-messenger observations.

\section*{Acknowledgments}
We thank  Miguel \'Angel Aloy and Martin Obergaulinger  for insightful discussions,  kindly sharing the data of the magnetorotational collapse model  employed  in this work, and comments on the manuscript. We also thank  Daniele Montanino for helpful comments on the manuscript. M.M.~gratefully acknowledges the hospitality of the Particle Astrophysics (AstroNu) group at the Niels Bohr Institute, whose stimulating environment significantly  contributed to the development of this work. M.M.~acknowledges financial support from the Scuola Superiore ISUFI (Universit\`a del Salento) through ``ISUFI - Ricerca \& Innovazione'' international mobility fellowship and from the research project TAsP (Theoretical Astroparticle Physics) funded by the Istituto Nazionale di Fisica Nucleare (INFN). In Copenhagen, this project has received support from the European Union (ERC, ANET, Project No.~101087058), the Villum Foundation (Project No.~13164),  and the Elite Research Prize  from the Danish Ministry of Higher Education and \hbox{Science} (Project No.~3142-00074B).
Views and opinions expressed are those of the authors only and do not necessarily reflect those of the European Union or the European Research Council. Neither the European Union nor the granting authority can be held responsible for them.

\bibliography{bibiliography}
\bibliographystyle{iopart-num}

\end{document}